\documentclass[10pt]{article}
\usepackage{authblk,mdframed}
\usepackage[top=2cm, bottom=2cm, left=2cm, right=2cm]{geometry}
\usepackage[T1]{fontenc} 
\usepackage{amssymb,amsmath,mathrsfs}
\usepackage{graphicx,rotate,subfigure}
\usepackage{cite}
\usepackage[colorlinks=true,
            linkcolor=red,
            urlcolor=blue,
            citecolor=blue]{hyperref}
\usepackage{color,xcolor}

\definecolor{ttl}{HTML}{31414D}
\definecolor{aut}{HTML}{610A0D}
\definecolor{abs}{HTML}{E0E0E0}
    \catcode`\_=11\relax
    \newcommand\email[1]{\_email #1\q_nil}
    \def\_email#1@#2\q_nil{%
      \href{mailto:#1@#2}{{\emailfont #1\emailampersat #2}}
    }
    \newcommand\emailfont{\ttfamily}
    \newcommand\emailampersat{{\color{blue}\small@}}
    \catcode`\_=8\relax  
\renewcommand\abstractname{\textbf{Abstract:}}
\renewenvironment{abstract}{%
\hfill\begin{mdframed}[backgroundcolor=abs]
\item[\hskip\labelsep\rmfamily\scshape\abstractname]}
{\par\noindent\end{mdframed}}
\makeatletter
\renewcommand\@maketitle{%
\hfill
\begin{minipage}{0.95\textwidth}
\vskip 2em
\let\footnote\thanks 
{\color{ttl} \LARGE \fontfamily{pnc}\selectfont \@title \par }
\vskip 1.5em
{\color{aut} \large \@author \par}
\end{minipage}
\vskip 1em \par
}

\let\TiTle=\title
\def\titlewithnum[#1]#2{\TiTle{{\rightline{\normalsize SINP/TNP/2015/#1} 
#2}}}
\def\titlenonum#1{\TiTle{#1}}
\def\title{\@ifnextchar[{\titlewithnum}{\titlenonum}}

\makeatother

\long\def\rpl#1!!#2!!{\textcolor{red}{#1} \textcolor{blue}{#2}}

\def\l{\lambda}
\def\ml{\mathscr{L}}
\def\tb{\tan\beta}
\def \order(#1){{\cal O} \left(#1 \right)}

\evensidemargin=\oddsidemargin
\def\Eqn#1{Eq.\ (\ref{#1})}

\allowdisplaybreaks
\begin{document}
%
%
\title[01]{New limits on $\mathbf{\tan\beta}$ for 2HDMs with $\mathbf{Z_2}$ symmetry}
\author[]{\fontfamily{phv}\selectfont  Dipankar Das\thanks{\email{d.das@saha.ac.in}}}
\affil[]{\it Saha Institute of Nuclear Physics, 1/AF Bidhan Nagar, Kolkata 700064, India}
%
\maketitle

\begin{abstract}
In two-Higgs-doublet models with exact $Z_2$ symmetry,
putting $m_h \simeq 125$ GeV at the alignment limit, the following limits on the heavy scalar
masses are obtained from the conditions of unitarity and stability of
the scalar potential: $m_H,~m_A,~m_{H^+} < 1$ TeV and $1/8 <
\tan\beta <8$.  The constraints from $b \to s \gamma$ and neutral meson mass differences, when superimposed on the unitarity constraints, put a tighter lower limit on $\tan\beta$ depending on $m_{H^+}$. It has also been shown that larger values of $\tb$ can be allowed by introducing soft breaking term in the potential at the expense of a correlation between $m_H$ and the soft breaking parameter.
\end{abstract}


\section{Introduction} \label{intro}
Extension of the Standard Model (SM) scalar sector is a common practice in constructing new physics models to address the shortcomings of the SM. Two-Higgs-doublet models (2HDMs)\cite{Branco:2011iw} are amongst the simplest of extensions that add only one extra $SU(2)$ doublet to the SM scalar sector. The tree level value of the  electroweak $\rho$-parameter remains unity for these types of extensions. In a general 2HDM, however, both the doublets ($\Phi_1$ and $\Phi_2$) can couple to each type of fermions. Consequently there will be two Yukawa matrices which, in general, are not diagonalizable simultaneously. This will introduce new flavor changing neutral currents (FCNC) mediated by neutral Higgses. It was shown by Glashow and Weinberg\cite{Glashow:1976nt} and independently by Paschos\cite{Paschos:1976ay} that Higgs mediated FCNC can be avoided altogether if fermions of a particular charge get their masses from the vacuum expectation value (vev) of a single scalar doublet.  This prescription was realized by employing a $Z_2$ symmetry under which one of the doublet is odd. Then there are four different possibilities for assigning $Z_2$ parities to the fermions so that Glashow-Weinberg-Pascos theorem is satisfied.  Following the usual convention, we shall always call $\Phi_2$ the doublet which couples to the up-type quarks. This leads to the following four types of 2HDMs:
\begin{itemize}
\item Type I: all quarks and leptons couple to only one scalar doublet $\Phi_2$ ;
\item Type II: $\Phi_2$ couples to up-type quarks, while $\Phi_1$ couples to down-type quarks and charged leptons (minimal supersymmetry conforms to this category);
\item Type X or lepton specific: $\Phi_2$ couples to all quarks, while $\Phi_1$ couples to all leptons;
\item Type Y or flipped: $\Phi_2$ couples to up-type quarks and leptons, while $\Phi_1$ couples to down-type quarks.
\end{itemize}

Now that a light Higgs boson has been observed by the ATLAS and CMS
Collaborations of the CERN Large Hadron Collider (LHC), the time is appropriate to revisit the
constraints on 2HDM parameter space arising
from the requirement of perturbative unitarity associated with the
scattering amplitudes together with the complementary constraints
coming from the flavor observables. Requirement of unitarity is a
consequence of probability conservation at the quantum level. Essentially, one takes
tree level amplitudes of appropriate sets of scattering processes, and
impose the unitarity condition that the $s$-wave scattering amplitude
$|a_0| < 1$. Equivalence theorem allows us to make the calculations easier by replacing the longitudinal gauge with the corresponding Goldstone bosons. 
In the context of the SM, Lee, Quigg and Thacker (LQT) had put an upper bound on the Higgs
boson mass, $m_h < m_{\rm LQT} = (8 \pi \sqrt{2}/3 G_F)^{1/2} \sim 1$
TeV\cite{Lee:1977eg}. The LQT-type analyses were later carried out to constrain the
nonstandard parameter spaces for several extensions of the SM scalar
sector. Detailed analyses of these constraints in the 2HDM context 
 already exist in the literature \cite{Maalampi:1991fb, Kanemura:1993hm, Akeroyd:2000wc,
   Horejsi:2005da,Gorczyca:2011he,Swiezewska:2012ej, Chakrabarty:2014aya}. In this
paper, we study the consequences of adding two additional inputs to
the existing analyses: ($i$) we use $m_h \simeq$ 125~GeV as input, which was an
unknown parameter earlier, and also note that the fitted values of the couplings of
the Higgs boson from the observed signal strengths into various fermion and
gauge boson channels show close conformity to the {\em alignment
  limit}, and ($ii$) the
  restrictions coming from two crucial flavor observables, namely,
  branching ratio of $b \to s \gamma$ and the meson mass splittings
 have been superimposed on the unitarity constraints . We observe that for a large class of 2HDM with {\em
  exact} $Z_2$ symmetry, the unitarity and flavor constraints {\em
  together} allow a rather restricted zone for $m_{H^+}$ and
$\tan\beta$, thus imposing {\em new} constraints on these parameters.
The limits on $\tan\beta$ are now independent of any other parameter,
its upper limit coming from unitarity and the lower limit from flavor
observables. We study these constraints in all four types of 2HDMs mentioned earlier. The limits are the strongest when the $Z_2$ symmetry is exact, while they get diluted by the soft symmetry breaking terms. We also
comment on what happens when instead of $Z_2$, softly broken
$U(1)$ symmetry is considered.

This paper is organized as follows: in Section~\ref{potential} we discuss the 2HDM scalar potential and the relevance of the alignment limit. Section~\ref{constraints} is divided into two parts. In the first part, we discuss the constraints arising from the requirements of unitarity and stability. Conclusions obtained from this part are independent of the Yukawa structure of the model. In the second part, we revisit the Yukawa sector dependent constraints originating from flavor data and superimpose the result on the unitarity constraints. Finally, the important findings are summarized in Section~\ref{conclusions}.

\section{The scalar potential} \label{potential}
The general scalar potential of a 2HDM invariant under a $Z_2$ ($\Phi_2\to -\Phi_2$) symmetry can be written as\cite{Gunion:1989we}
\begin{eqnarray}
 V &=& 
 \lambda_1 \left( \Phi_1^\dagger\Phi_1 - \frac{v_1^2}{2} \right)^2 
+\lambda_2 \left( \Phi_2^\dagger\Phi_2 - \frac{v_2^2}{2} \right)^2 
 +\lambda_3 \left( \Phi_1^\dagger\Phi_1 + \Phi_2^{\dagger}\Phi_2 
- \frac{v_1^2+v_2^2}{2} \right)^2
\nonumber \\
&& +\lambda_4 \left(
(\Phi_1^{\dagger}\Phi_1) (\Phi_2^{\dagger}\Phi_2) -
(\Phi_1^{\dagger}\Phi_2) (\Phi_2^{\dagger}\Phi_1)
\right)  + \lambda_5 \left( {\rm Re}~  \Phi_1^\dagger\Phi_2  - \frac{v_1v_2}{2}  \right)^2 
+ \lambda_6 \left({\rm Im}~ \Phi_1^\dagger\Phi_2  \right)^2 \,,
\label{notation2}
\end{eqnarray}
where, the bilinear term proportional to $\lambda_5$  breaks the $Z_2$ symmetry softly. This soft breaking term, in its conventional parametrization, is written as $-m_{12}^2(\Phi_1^\dagger\Phi_2+\Phi_2^\dagger\Phi_1)$. The connection between $m_{12}^2$ and $\l_5$ is given by\cite{Bhattacharyya:2014oka}
\begin{equation}
m_{12}^2 = \frac{\l_5}{2}v_1v_2 \,.
\label{m12}
\end{equation}
Defining $\tan\beta \equiv v_2/v_1$ to be the ratio of the two vevs, we also remember that it is the combination $m_{12}^2/(\sin\beta\cos\beta)$, not $m_{12}^2$ itself, which controls the nonstandard masses\cite{Gunion:2002zf}. In view of these facts, $\l_5$, rather than $m_{12}^2$, constitutes a convenient parameter that can track down the effect of soft breaking.
Note that, unlike the {\em inert} case, \Eqn{notation2} implicitly assumes that the $Z_2$ symmetry is also broken spontaneously, {\it i.e.}, both the doublets receive vevs. In this article, we shall only consider 2HDMs where the value of $\tan\beta$ is nonzero and finite. We have also assumed that all the potential parameters are real, {\em i.e.}, CP symmetry is exact in the scalar potential. The last assumption allows us to define electrically neutral mass eigenstates which are also eigenstates of CP. Here there will be a total of five physical scalars: a pair of CP-even scalars ($h$ and $H$ with $m_H>m_h$), one CP-odd scalar ($A$) and a pair of charged scalars ($H^\pm$). For the transition into the mass basis, one needs to rotate the original fields in \Eqn{notation2} by an angle $\beta$ in the charged and the CP-odd sectors, whereas, the rotation angle involved in the CP-even sector is denoted by $\alpha$. The latter angle is defined through the relation
\begin{eqnarray}
\tan 2\alpha = \frac{2\left(\lambda_3+\frac{\lambda_5}{4} \right)v_1v_2}{\lambda_1v_1^2- \lambda_2v_2^2+\left(\lambda_3+\frac{\lambda_5}{4}\right)(v_1^2-v_2^2)} \,.
\end{eqnarray}

Note that there were eight parameters to start with: $v_1,~v_2$ and 6 lambdas. One can trade $v_1$ and $v_2$ for $v=\sqrt{v_1^2+v_2^2}$ and $\tan\beta$. All the lambdas except $\lambda_5$ may be traded for 4 physical Higgs masses and $\alpha$. The relations between these two equivalent sets of parameters are given below~:
\begin{subequations}
\begin{eqnarray}
\lambda_1 &=& \frac{1}{2v^2\cos^2\beta}\left[m_H^2\cos^2\alpha  +m_h^2\sin^2\alpha  -\frac{\sin\alpha\cos\alpha}{\tan\beta}\left(m_H^2-m_h^2\right)\right] -\frac{\lambda_5}{4}\left(\tan^2\beta-1\right) \,, \\
\lambda_2 &=& \frac{1}{2v^2\sin^2\beta}\left[m_h^2\cos^2\alpha  +m_H^2\sin^2\alpha  -\sin\alpha\cos\alpha\tan\beta\left(m_H^2-m_h^2\right) \right] -\frac{\lambda_5}{4}\left(\cot^2\beta-1\right) \,, \\
\lambda_3 &=& \frac{1}{2v^2} \frac{\sin\alpha\cos\alpha}{\sin\beta\cos\beta} \left(m_H^2-m_h^2\right) -\frac{\lambda_5}{4} \,, \\
\lambda_4 &=& \frac{2}{v^2} m_{H^+}^2 \,, \\
\lambda_6 &=& \frac{2}{v^2} m_A^2 \,.
\end{eqnarray}
\label{inv2HDM}
\end{subequations}
Among these, $v=246$ GeV is already known and if it is assumed that the lightest CP-even Higgs is what has been observed at the LHC, then $m_h=125$ GeV is also known. The rest of the parameters need to be constrained from theoretical as well as experimental considerations.

The first simplification occurs if one keeps in mind that the experimental values of the Higgs signal strengths into different decay channels are increasingly leaning towards the corresponding SM predictions\cite{Atlas:signal,cms:signal}. In the 2HDM context, this implies the {\em alignment} condition\cite{Gunion:2002zf}
\begin{eqnarray}
\sin(\beta-\alpha) \approx 1 \,,
\label{alignment}
\end{eqnarray}
which means, $h$ will have the exact same tree-level couplings with the vector bosons and fermions as in the SM. In view of the recent global fits for 2HDMs using the LHC Higgs data, \Eqn{alignment} is a reasonable assumption\cite{Eberhardt:2013uba,Coleppa:2013dya, Chen:2013rba, Craig:2013hca, Dumont:2014wha, Bernon:2014vta}.

Next, one has to ensure that there should not exist any direction in the field space along which the potential of \Eqn{notation2}
becomes infinitely negative, {\it i.e.}, the potential is bounded from below. The necessary and sufficient conditions for this can be found to be\cite{Klimenko:1984qx, Maniatis:2006fs}
  \begin{subequations}
  \label{stability}
  \begin{eqnarray}
&&  \lambda_1+\lambda_3 >0 \,, \\
&&  \lambda_2+\lambda_3 >0 \,, \\
&& 2\lambda_3+\lambda_4 +2\sqrt{(\lambda_1+\lambda_3)(\lambda_2+\lambda_3)} >0 \,, \\
&& 2\lambda_3+\frac{\lambda_5+\lambda_6}{2} -\frac{|\lambda_5-\lambda_6|}{2}+2\sqrt{(\lambda_1+\lambda_3)(\lambda_2+\lambda_3)} >0 \,.
  \end{eqnarray}
  \end{subequations}

To obtain the constraints from tree-unitarity, we construct an $S$-matrix using different two-body states to label its different rows and columns. The $\ell=0$ partial wave amplitudes for different $2\to 2$ scattering processes constitute the elements of this $S$-matrix. The explicit expressions for the eigenvalues of this matrix are listed below\cite{Maalampi:1991fb, Kanemura:1993hm, Akeroyd:2000wc, Horejsi:2005da}:
 \begin{subequations}
 \begin{eqnarray}
 a_1^\pm &=& 3(\l_1+\l_2+2\l_3) \pm \sqrt{9(\l_1-\l_2)^2+ \left(4\l_3+\l_4+\frac{\l_5+\l_6}{2} \right)^2} \,, \\
 a_2^\pm &=& (\l_1+\l_2+2\l_3) \pm \sqrt{(\l_1-\l_2)^2+\frac{1}{4} \left(2\l_4-\l_5-\l_6 \right)^2} \,, \\
 a_3^\pm &=& (\l_1+\l_2+2\l_3) \pm \sqrt{(\l_1-\l_2)^2+\frac{1}{4} \left(\l_5-\l_6 \right)^2} \,, \\
 b_1 &=& 2\l_3-\l_4-\frac{1}{2}\l_5+\frac{5}{2}\l_6 \,, \\
 b_2 &=& 2\l_3+\l_4-\frac{1}{2}\l_5+\frac{1}{2}\l_6 \,, \\
 b_3 &=& 2\l_3-\l_4+\frac{5}{2}\l_5-\frac{1}{2}\l_6 \,, \\
 b_4 &=& 2\l_3+\l_4+\frac{1}{2}\l_5-\frac{1}{2}\l_6 \,, \\
 b_5 &=& 2\l_3+\frac{1}{2}\l_5+\frac{1}{2}\l_6 \,, \\
 b_6 &=& 2(\l_3+\l_4)-\frac{1}{2}\l_5-\frac{1}{2}\l_6 \,.
 \end{eqnarray}
 \label{unieigenvalues}
 \end{subequations}
The requirement of tree-unitarity then restricts each of the above eigenvalues as
\begin{eqnarray}
|a_i^\pm|,~|b_i| \le 16\pi \,.
\label{unitarity}
\end{eqnarray}
Now it is the time to investigate how the constraints from unitarity and stability restrict the parameter space in the {\em alignment limit} defined by \Eqn{alignment}.
\begin{figure}
\centering
\includegraphics[scale=0.36]{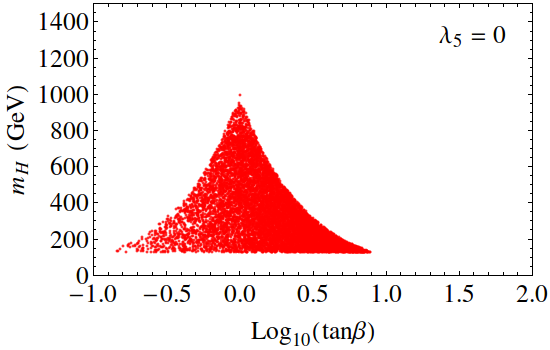}~~
\includegraphics[scale=0.36]{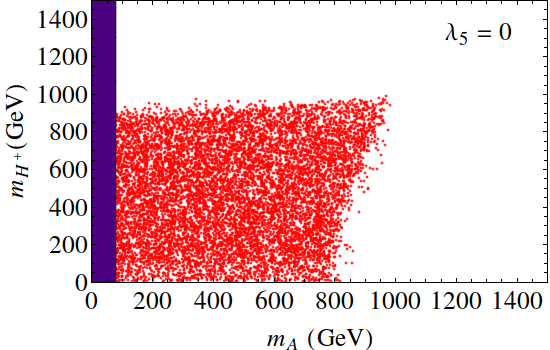}
\caption{\em Allowed region from unitarity and stability for exact $Z_2$ symmetry. In the right panel, the dark (purple) shaded region is excluded from direct search\cite{Searches:2001ac}.}
\label{f:uniZ2}
\end{figure}

\section{Constraints on the 2HDM parameter space} \label{constraints}
To begin with, we consider the case when the $Z_2$ symmetry is exact in the scalar potential {\em i.e.}, $\l_5=0$. For this, we have generated millions of random points in the $\{\tan\beta,~m_H,~m_A,~m_{H^+}\}$ space. The individual parameters have been varied in the following range:
\begin{eqnarray}
\tan\beta \in [0.1,100]\,, ~~ m_H\in [125,2000] \,, ~~m_A\in [0,2000] \,, ~~ m_{H^+}\in [0,2000] \,.
\end{eqnarray}
Those points which successfully negotiate the unitarity and stability bounds have been plotted in Fig.~\ref{f:uniZ2}. Some noteworthy features are listed below~:
\begin{itemize}
\item From the left panel, one can read the limit on $\tan\beta$, $1/8<\tan\beta<8$.
\item Strong correlation exists between the upper limit of $m_H$ and $\tan\beta$.
\item Limits on the masses are, $m_H,~m_A,~m_{H^+}<1$ TeV.
\end{itemize}
\begin{figure}
\centering
\includegraphics[scale=0.3]{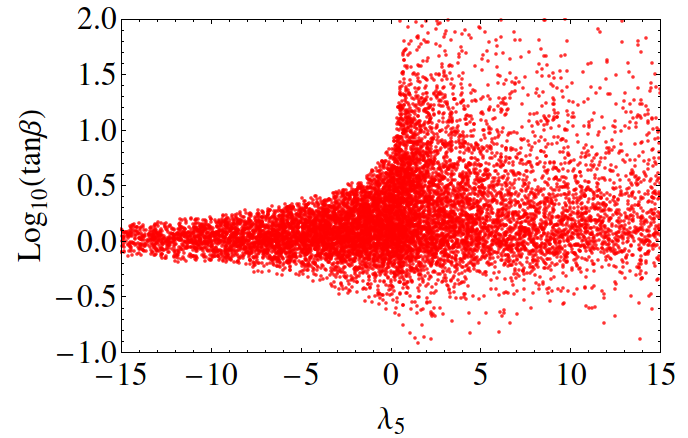}~~
\includegraphics[scale=0.3]{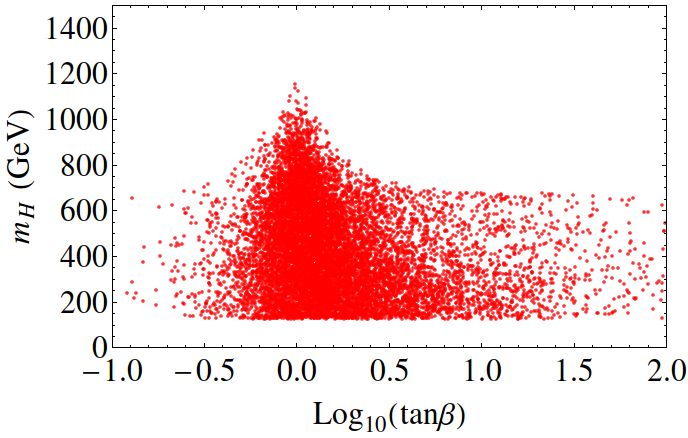}
\caption{\em Relaxation of the unitarity and stability constraints on $\tan\beta$ in the presence of soft breaking. In the right panel, the vertical width of the of the tail in the region where $\tb$ is much away from unity is caused by the variation of $\l_5$ in the range $[-15,15]$.}
\label{f:l5tb}
\end{figure}
The reason for the above features can be traced back to the eigenvalues of \Eqn{unieigenvalues}. First two constraints for boundedness in \Eqn{stability} can be combined into
\begin{eqnarray}
\l_1+\l_2+2\l_3 > 0 \,.
\end{eqnarray}
This, then together with the condition $|a_1^\pm| < 16\pi$, implies
\begin{eqnarray}
&& 0< \l_1+\l_2+2\l_3 < \frac{16\pi}{3} \,, \\
\Rightarrow && 0 < \left( m_H^2 -\frac{1}{2}\l_5v^2 \right)(\tan^2\beta+\cot^2\beta) +2m_h^2 <\frac{32\pi v^2}{3} \,,
\label{unitb}
\end{eqnarray}
where the last expression is obtained from the previous one by using \Eqn{inv2HDM} in the alignment limit. Keeping in mind that $m_H>125$ GeV, this will put a limit on $\tan\beta$ (as well as $\cot\beta$) when $\lambda_5=0$. Since the minimum value of ($\tan^2\beta+\cot^2\beta$) is 2  when $\tan\beta=1$, the maximum possible value of $m_H$ occurs at $\tan\beta=1$. In summary, the inequality (\ref{unitb}) explains the $\tan\beta$ dependent bound on $m_H$ as depicted in the left panel of Fig.~\ref{f:uniZ2}.
\begin{figure}
\includegraphics[scale=0.22]{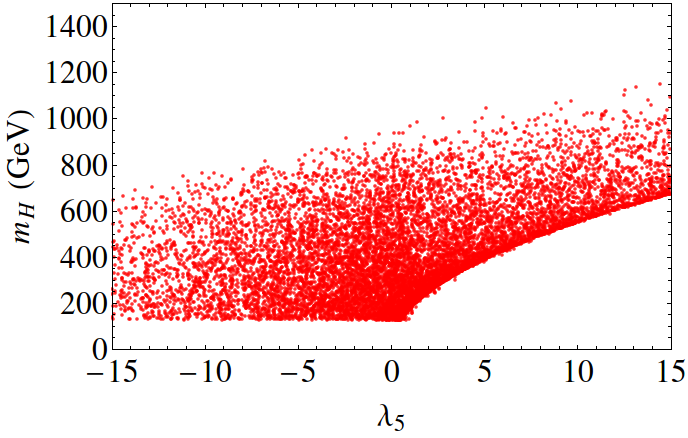}~~
\includegraphics[scale=0.22]{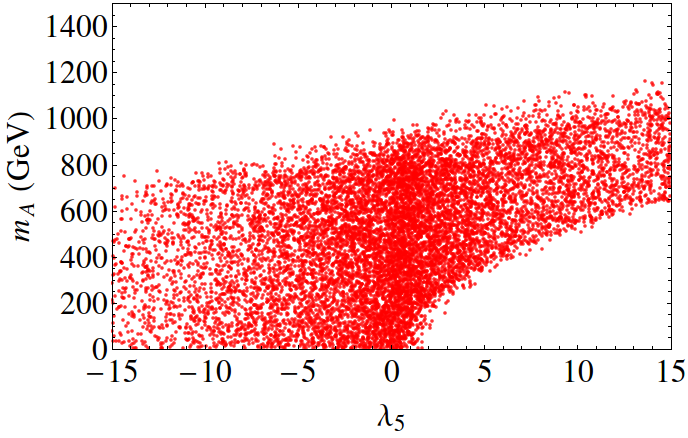}~~
\includegraphics[scale=0.22]{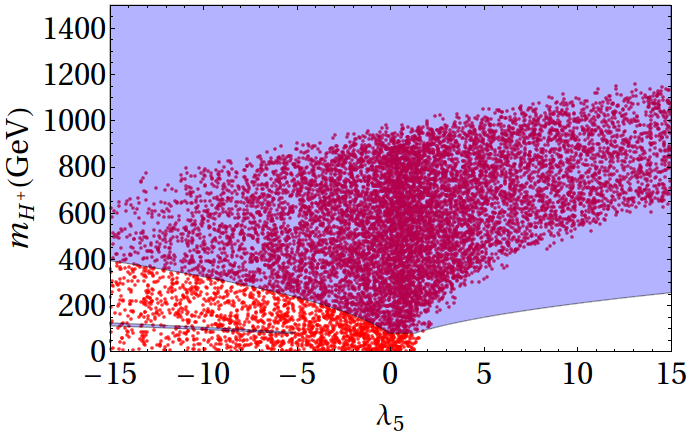}
\caption{\em Effect of soft breaking on the constraints on the nonstandard masses. The (light blue) shaded region in the rightmost panel represents the combined allowed region from direct search and the diphoton signal strength at 95\% C.L.}
\label{f:l5mass}
\end{figure}

\Eqn{unitb} also implies that the restriction on $\tan\beta$ will be lifted for $1/2\l_5 v^2 > (m_H^2)_{\rm min} = (125~{\rm GeV})^2$. Once this condition is satisfied, $m_H^2$ will have the chance to saturate to $1/2\l_5v^2$ making the difference between them to vanish in \Eqn{unitb}. In fact, to a very good approximation, one can use 
\begin{eqnarray}
m_H^2 \approx 1/2\l_5v^2
\label{corr}
\end{eqnarray}
 for $\tan\beta > 5$.

To understand the restrictions on $m_A$ and $m_{H^+}$, we use the triangle inequality to note the following~:
 \begin{subequations}
 \begin{eqnarray}
 |b_1-b_3| \equiv 3|\l_6-\l_5| < 32\pi \,, && \Rightarrow~ |m_A^2-\frac{1}{2}\l_5v^2| < \frac{16\pi v^2}{3} \,, \label{limma} \\
 |b_6-b_3| \equiv 3|\l_4-\l_5| < 32\pi \,, && \Rightarrow~ |m_{H^+}^2-\frac{1}{2}\l_5v^2| < \frac{16\pi v^2}{3} \,.
 \label{limm1+}
 \end{eqnarray}
 \end{subequations}
Because of Eqs.~(\ref{limma}) and (\ref{limm1+}) one expects to put limits on $m_A$ and $m_{H^+}$ respectively, when $\l_5=0$. Additionally, note that due to the inequality 
\begin{eqnarray}
|b_1-b_6| \equiv 3|\l_6-\l_4| < 32\pi \,, && \Rightarrow~ |m_A^2-m_{H^+}^2| < \frac{16\pi v^2}{3} \,,
\label{diff}
\end{eqnarray}
it is expected that the splitting between $m_A$ and $m_{H^+}$ will be {\em always} restricted in a 2HDM. It is also interesting to note that the conclusions obtained from Eqs.~(\ref{limma}), (\ref{limm1+}) and (\ref{diff}) {\em do not} depend on the imposition of the alignment condition.

Next we shall investigate the implications of the soft breaking parameter on these constraints. We have varied $\l_5$ in the range $[-15,15]$ for this purpose. From \Eqn{unitb} one can observe that the space for $\tan\beta$ is squeezed further if $\l_5 <0$ but the bound is relaxed if $\l_5>0$. This feature emerges from the left panel of Fig.~\ref{f:l5tb}. One can also see from \Eqn{unitb} that $m_H^2$ must follow $1/2\l_5v^2$ if $\tan\beta$ moderately deviates from unity. This feature is reflected by the horizontal tail in the right panel of Fig.~\ref{f:l5tb} on both sides of the peak. The vertical width of the tail is caused by the variation of $\l_5$ in the range $[-15, 15]$. On the other hand, from Eqs.~(\ref{limma}), (\ref{limm1+}) and (\ref{unitb}), it should be noted that the upper bounds on the nonstandard scalar masses will be relaxed for $\l_5>0$ but will get tighter for $\l_5<0$. Fig.~\ref{f:l5mass} reflects these features where one can see that this dependence is rather weak. It is worth mentioning at this point that if one uses a softly broken $U(1)$ symmetry instead of the usual $Z_2$ symmetry, the soft breaking parameter gets related to the pseudoscalar mass as $m_A^2=1/2\l_5v^2$. Consequently, the correlation between $m_H$ and $\l_5$ in the leftmost panel of Fig.~\ref{f:l5mass} transforms into the degeneracy between $m_H$ and $m_A$. Detailed analysis of the scalar sector, for the softly broken $U(1)$ scenario, has been carried out in \cite{Bhattacharyya:2013rya,Biswas:2014uba}.

It is also important to note that the production as well as the tree-level decay widths of $h$ remain unaltered from the corresponding SM expectations due to the imposition of alignment limit of \Eqn{alignment}. But the loop induced decay modes of $h$, such as $h\to\gamma\gamma$ and $h\to Z\gamma$, will pick up additional contributions due the presence of the charged scalar in loops. For example, the diphoton signal strength ($\mu_{\gamma\gamma}$), in general, depends on both $\l_5$ and $m_{H^+}$\cite{Bhattacharyya:2014oka}. The current measurement by CMS gives $\mu_{\gamma\gamma}=1.14^{+0.26}_{-0.23}$\cite{Khachatryan:2014ira}, whereas ATLAS measures $\mu_{\gamma\gamma}$ to be $1.17\pm 0.27$\cite{Aad:2014eha}. In addition to this, the direct search limit of $m_{H^+}>80$~GeV\cite{Searches:2001ac} should also be taken into account. Considering all of these experimental constraints, the allowed region at 95\% C.L. has been shaded (in light blue) in the rightmost panel of Fig.~\ref{f:l5mass}. Only those points that lie within the shaded region survive both the theoretical and experimental constraints.

\subsection{Yukawa sector and flavor constraints}
\begin{figure}
\begin{center}
\includegraphics[scale=0.4]{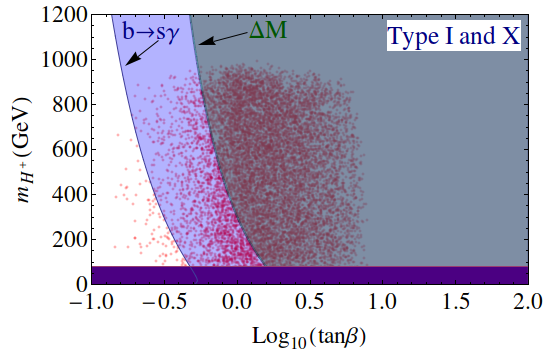} ~~
\includegraphics[scale=0.4]{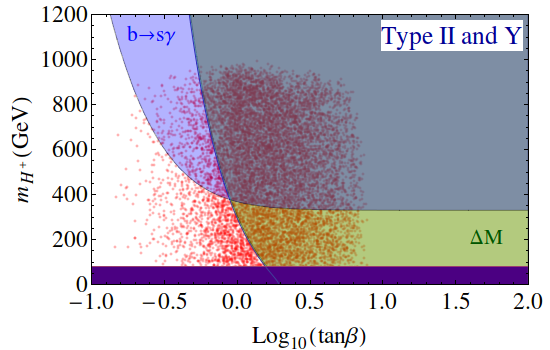}
\end{center} 
\caption{\em Constraints on $\tan\beta$ and the charged Higgs mass
  from unitarity and flavor physics. The left panel corresponds to
  Type~I and X and the right panel to Type~II and Y scenarios. The lower
  horizontal dark (purple) strip in both the panels corresponds to the direct search
  limit of 80~GeV\cite{Searches:2001ac}. The ligther shades represent allowed regions from individual flavor observables. The scattered points are allowed from unitarity and stability for 2HDMs with exact $Z_2$ symmetry.}
\label{flavor}
\end{figure}
Now we shall concentrate on the constraints on the charged scalar mass ($m_{H^+}$), imposed by the measured values of $b\to s\gamma$ branching ratio\cite{Amhis:2012bh} and neutral meson mass differences ($\Delta M$)\cite{Agashe:2014kda}. Since we are concerned with the quark sector only, the constraints will be the same for Type~I and Type~X models. The same is true for Type~II and Type Y models. In the following, we spell out the relevant parts of the charged scalar Yukawa interaction:
 \begin{subequations}
 \label{yuk}
 \begin{eqnarray}
\ml_{H^+}^{(\rm I~or~X)} &=& \left[\frac{\sqrt{2}H^+}{v}\cot\beta  \Big\{\bar{u}_R\left(D_uV\right)d_L-\bar{u}_L\left(VD_d\right)d_R \Big\} +{\rm h.c.} \right] \,,
\label{1y} \\
\ml_{H^+}^{(\rm II~or~Y)} &=& \left[\frac{\sqrt{2}H^+}{v} \Big\{\cot\beta \bar{u}_R\left(D_uV\right)d_L +\tan\beta\bar{u}_L\left(VD_d\right)d_R \Big\} +{\rm h.c.} \right] \,,
\label{2y}
 \end{eqnarray}
 \end{subequations}
where, $V$ is the CKM matrix and $D_{u,d}$ are diagonal mass matrices in the up- and down-quark sectors respectively. In writing \Eqn{yuk}, we have suppressed the flavor indices. Thus $u$ and $d$ should be interpreted as three element column matrices.

For the process $b\to s\gamma$, the major new physics contributions come from charged scalar loops. We have added the new physics contribution to the SM one at the amplitude level and therefore have taken the interference into account. The branching ratio is then compared with the experimental value, $(3.55\pm 0.26)\times 10^{-4}$\cite{Amhis:2012bh}, to obtain the allowed region at 95\% C.L. in Fig.~\ref{flavor}.
As can be seen from \Eqn{2y}, for Type~II and Y models, in the charged Higgs Yukawa interaction, the up-type Yukawa coupling is multiplied by $\cot\beta$ while the down-type Yukawa is multiplied by $\tan\beta$. Their product is responsible for setting $\tan\beta$-independent limit\footnote{Recently this bound has been changed to $m_{H^+}>480$~GeV using the NNLO result along with the updated experimental value\cite{Misiak:2015xwa}. This means that the $b\to s\gamma$ boundary on the right panel of Fig.~\ref{flavor} will be shifted upwards slightly. But this will hardly modify the lower bound on $\tb$ which mainly comes from $\Delta M$.} $m_{H^+}>320$ GeV for $\tan\beta >1$\cite{WahabElKaffas:2007xd,Mahmoudi:2009zx,Deschamps:2009rh}. This feature has been depicted in the right panel of Fig.~\ref{flavor}. In Type~I and X models, on the other hand, each of these couplings picks up a $\cot\beta$ factor. This is why there is essentially no bound on $m_{H^+}$ for $\tan\beta >1$ in these models\cite{Mahmoudi:2009zx}. This character of Type~I and X models emerges from the left panel of Fig.~\ref{flavor}.

The dominant new physics contributions to neutral meson mass differences come from the charged scalar box diagrams. Note that, the constraint arising from $\Delta M$ in the $m_{H^+}$-$\tb$ plane, is slightly stronger than that from the precision measurement of the $Z\to b\bar{b}$ branching ratio\cite{Deschamps:2009rh}.  In Fig.~\ref{flavor}, allowed regions have been shaded assuming that the new physics contributions saturate the experimental values of $\Delta M$\cite{Agashe:2014kda}. Since the amplitudes for the new box diagrams receive prevailing contributions from the up-type quark masses which, for all four variants of 2HDMs, comes with a $\cot\beta$ prefactor, the overall charged scalar contribution to the amplitude goes as $\cot^4\beta$ due to the presence of four charged scalar vertices in the box diagram. Not surprisingly, $\Delta M$ offers a stronger constraint than $b\to s\gamma$ for $\tan\beta <1$ because, in this region, the new physics amplitude for the latter goes as $\cot^2\beta$.

Things become more interesting when the above flavor constraints are superimposed on top of the constraints from unitarity and stability. Most stringent constraints are obtained when $Z_2$ symmetry is exact in the scalar potential, {\it i.e.}, $\l_5=0$. In Fig.~\ref{flavor}, the scattered points span the region allowed by the combined constraints of unitarity and stability for the case of exact $Z_2$ symmetry. Only those points which lie within the common shaded region survive when all the constraints are imposed. For 2HDMs of all four types, one can read the bound on $\tb$ as
\begin{eqnarray}
0.5 < \tb <8 \,.
\end{eqnarray}
However, in order to allow a lighter charged scalar in the ballpark of 400~GeV or below, one must require $1<\tb<8$. It should be remembered that, the lower bound on $\tb$ mainly comes from the flavor data, whereas the upper limit, for the case of exact $Z_2$ symmetry, is dictated by unitarity and stability. In the presence of a soft breaking parameter, the upper bound will be lifted allowing $\tb$ to take much larger values at the expense of a strong correlation between the soft breaking parameter and $m_H$ as depicted by \Eqn{corr}.

\section{Conclusions} \label{conclusions}
In this paper, we have revisited the constraints from tree-unitarity and stability in the context of 2HDMs. The observed scalar at LHC has been identified with the lightest CP-even scalar of the model. The {\em alignment limit} has been imposed in view of the conformity of the LHC Higgs data with the SM predictions. These are the {\em new} informations that became available only after the Higgs discovery. If the $Z_2$ symmetry is exact in the potential, it is found that all the nonstandard masses are restricted below 1~TeV from unitarity with the upper limit on $m_H$ being highly correlated to $\tb$. The value of $\tb$ is also confined within the range $1/8<\tb<8$ from unitarity and stability. The constraints from flavor data severely restrict the region with $\tb<1$. Therefore, for an exact $Z_2$ symmetry, $\tb$ is bounded within a very narrow range of $1<\tb<8$ when a light charged scalar with mass around 400~GeV is looked for.

In the presence of an appropriate soft breaking parameter the upper bound on $\tb$ will be diluted. However, for large values of $\tb$, the unitarity and stability conditions will render a strong correlation between the soft breaking parameter and $m_H$ as appears in \Eqn{corr}. It is also worth noting that the value of $\mu_{\gamma\gamma}$ can play a crucial role in the presence of soft breaking. For example, if $\mu_{\gamma\gamma}$ is measured to be consistent with the SM expectation with 5\% accuracy then one can conclude $m_{H^+}^2 \approx 1/2\l_5v^2$\cite{Bhattacharyya:2014oka} for any value of $\tb$. Thus, for large values of $\tb$ one may expect $m_H^2\approx m_{H^+}^2\approx 1/2\l_5v^2$. In this limit, the heavier
nonstandard scalars truly decouple from the low energy observables. Thus, as has been emphasized in\cite{Bhattacharyya:2014oka}, proper decoupling of the nonstandard scalars
necessitates the presence of a soft breaking term in the scalar potential.

To sum up, when flavor constraints are superimposed on the constraints from unitarity and stability, the value of $\tb$ is restricted within a very narrow range of $1<\tb<8$ for 2HDMs with exact $Z_2$ symmetry. Larger values of $\tb$ can be allowed by introducing suitable soft breaking parameter in the scalar potential, but the theoretical and experimental constraints impose certain correlations between nonstandard masses and the soft breaking parameter. This makes the theory much more predictive in the large $\tb$ region.

\paragraph*{Acknowledgements:} I thank G.~Bhattacharyya for many helpful discussions during different stages of this work. I also thank Department of Atomic Energy, India for financial support.


\bibliographystyle{JHEP}
\bibliography{ref_Z2.bib}
\end{document}